# DETC2011

# COMPLEX EIGENVALUE ANALYSIS FOR STRUCTURES WITH VISCOELASTIC BEHAVIOR.

**G. Chevallier**
LISMMA EA 2336
Supméca
3 rue Fernand HAINAUT
93407 SAINT-OUEN
FRANCE

**F. Renaud**
Supméca
93400 SAINT-OUEN

**J.-L. Dion**
Supméca
93400 SAINT-OUEN

**ABSTRACT**
This document deals with a method for eigenvalue extraction for the analysis of structures with viscoelastic materials. A generalized Maxwell model is used to model linear viscoelasticity. Such kind of model necessitates a state-space formulation to perform eigenvalue analysis with standard solvers. This formulation is very close to ADF formulation. The use of several materials on the same structure and during the same analysis may lead to a large number of internal states. This article purpose is to identify simultaneously all the viscoelastic materials and to constrain them to have the same time-constants. As it is usually possible, the size of the state-space problem is therefore widely reduced. Moreover, an accurate method for reducing mass and stiffness operators is proposed; The enhancement of the modal basis allows to obtain good results with large reduction. As the length of the paper is limited, only theoretical development are presented in the present paper while numerical results will be presented in the conference.

## INTRODUCTION
Many mechanical systems are damped with viscoelastic materials. This helps to obtain highly damped structures and thus to limit the vibration levels. Although the viscoelastic behavior of materials is of great importance in order to obtain accurate eigenvalues and eigenmodes, the assumption of purely elastic materials is very commonplace for frequency analysis with Finite Element (FE) models. In order to carry out realistic Complex Eigenvalues Analysis (CEA) in dynamics, one needs to measure, identify and to model the viscolelastic behavior of the structure;

## Theoretical aspects
Viscoelastic behavior may be described using internal states in the time domain, see [1] or [2], or rational fraction with poles and zeros in the frequency domain. This way had led to the famous GHM [1] and ADF [2] models useful both in time and frequency domain. These aspects are well summarized in the paper of Vasques& al. [3]. The rheological model associated with such kind of models is the well known Generalized Maxwell Model, see [4] , [5]or [6] for examples of the use of this model in its rheological form. The use of fractional derivative models with rational exponents leads also to model visco elasticity with internal states, see sorrentino& al. [7].

A particular focus on CEA is done in the present paper because the analysis of the stability of non-linear systems is often performed using this tool :Aeroelasticity or friction induced vibrations are studied with CEA, see [8] for example. Adding viscoelastic materials in these simulations leads to augment the order of the eigenvalue problem to solve. Several methods had been developed to solve this kind of problem from non linear eigenvalue extraction, see Daya& al. [9] for example, to the use of state space formulation. But these two approaches are quite time-consuming in terms of CPU use.

**This paper aims to give an original state space formulation which allows to reduce the computation cost of CEA.** The main originalities of this paper are the use of common time-relaxation constants for all the viscoelastic materials that are used in the simulations and the use of a enhanced modal basis for the projections of mass and stiffness FE operators. To achieve this goal, identification and modeling computations have to be linked in order to obtain common time-constants. The identification process used in this paper is well described in the paper of Renaud & al. [6].

The first part of this paper deals with the use of Generalized Maxwell Models in Finite Element Formulation. As the authors are mainly interested by Complex Eigenvalue Analysis, they propose a state-space formulation of the equations. Solving the problem obtained leads to the determination of complex modes and eigenvalues. This kind of formulation is very close to the formulation ADF proposed by Lesieutre. Nevertheless, the use of several viscoelastic materials can lead to determination of a large number of internal states and thus to very large state space model. Thus, according to the idea proposed by Trindade& al. [10], the authors propose to use a modal basis projection to reduce the size of the mass and stiffness operators.

In the second part of the paper, the authors propose to identify simultaneously the parameters of the Generalized Maxwell Model of all the materials under the constraint that they must have the same time constants. This assumption allows limiting the number of internal state in the state space formulation and thus reducing the computation time. Many numerical results are given on the following example to illustrate the paper.

### Numerical Example

In order to illustrate the analytical developments and to demonstrate the computing efficiency, the authors present a simple example of multi-layered beam with three different viscoelastic materials, see figure 1. The beam is clamped on the left and free on the right. It has been designed to have its first natural frequency at 70 Hz. Moreover, the authors are interested in the first 15 modes only. These modes have been computed using the Finite Element Method in the industrial code ABAQUS. The structure has been meshed with C3D8 Linear Volumetric Elements. Detailed results will be presented in the conference.

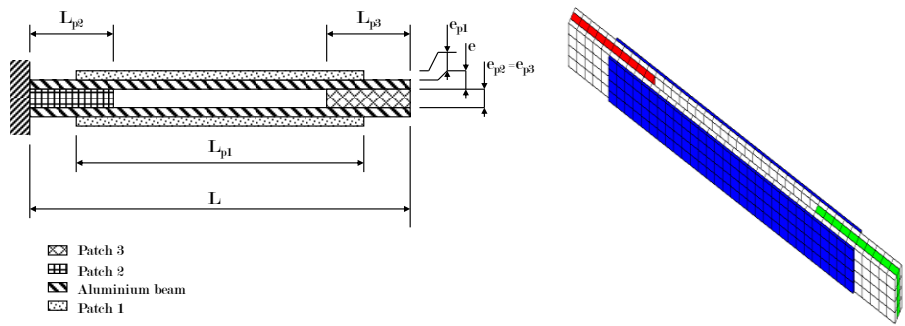

**Figure 1.** Left : Schema of the multilayered beam. Dimensions are : L=0.2m – Lp1=0.15m – Lp2=0.05m – Lp3=0.05m– e=0.002m– ep1=0.001m– ep2= ep3=0.003m – depth for all the layers : d=0.02m. Patch are made with polymer materials. Beam is constituted with aluminum thin layers.
Right : View of the meshed beam : C3D8 Elements – Red : Patch 2 – Green : Patch 3 – White : Aluminium layers.

An identification procedure has been used to determine the parameters of the viscoelastic models. The three materials are supposed to be frequency dependent and to behave linearly against the loads ; So that they are assumed to be well described with a complex Young Modulus : $E^*=E(1+j\eta)$. In this definition $E$ is often called "Storage Modulus" and $\eta$ is called "Loss Factor". In addition, the Phase is defined as follow : $\tan\phi=\eta$. Material 1 has a quite constant phase over the studied frequency range : i.e. [70... 4000] Hz. Material 2 has a linearly increasing phase over the studied frequency range from 5 deg. to 10 deg. Material 3 has a decreasing phase from 20 deg. to 15 deg.

The first natural modes of the structure have been computed considering the Long Term Moduli $E\infty$ of the materials. This allows to compute the strain energies of each viscoelastic parts and then to approximate the modal damping of the structure. This technic is widely used in mechanical engineering to perform a rapid evaluation of the damping of such kind of structure. In this paper, we will use it as a reference to be compared to more accurate technics.

These modes are computed using ABAQUS to build the mass and stiffness operators and to compute the first eigenmodes using Lanczos algorithm. These modes are mainly bending modes and few of them are torsionnal modes.

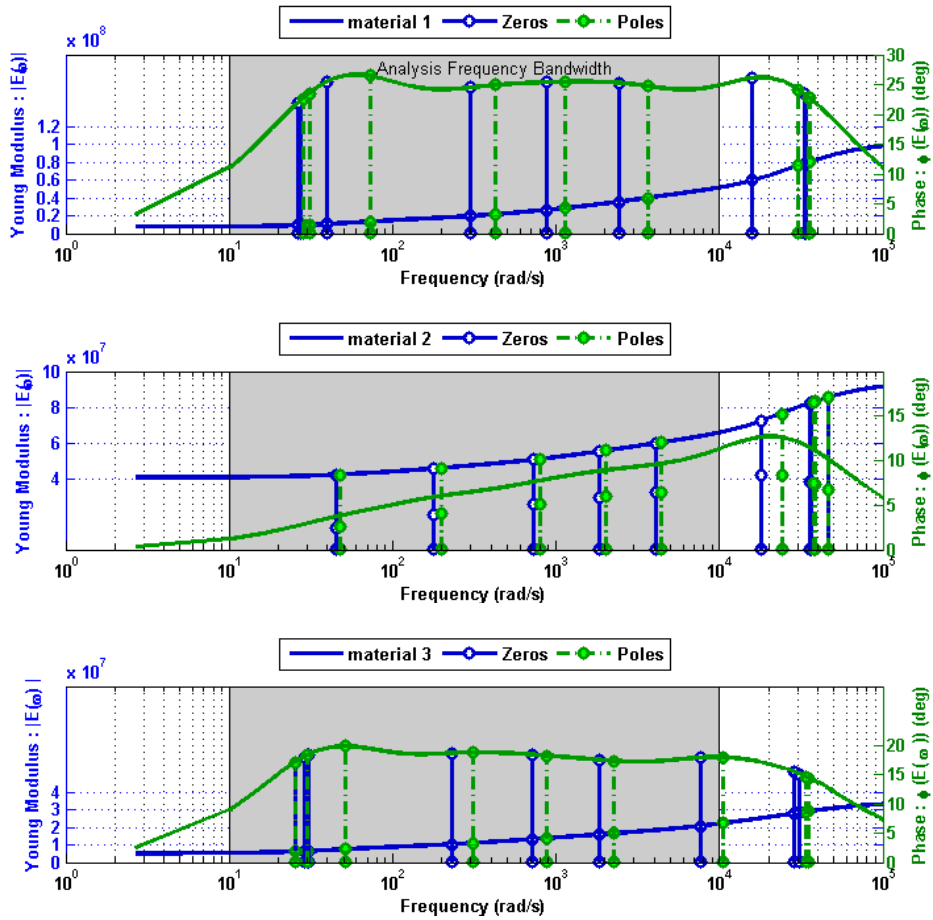

Figure 2. Frequency dependence of the three viscoelastic materials ; Frequency independent data are :
Material 1 : $E\infty$=10MPa - $v$=0.49 – $\rho$=1000 kg/m$^3$
Material 2 : $E\infty$=20MPa - $v$=0.49 – $\rho$=1000 kg/m$^3$
Material 3 : $E\infty$=5MPa - $v$=0.49 – $\rho$=1000 kg/m$^3$

**Aims of the present work**

The present work aims to compute complex eigenvalues in order to obtain with a great precision and quick computation: the natural frequencies and the modal damping. Due to the frequency dependence of the materials, the generalized eigenvalue problem is quite complex to compute. The authors' purpose is to reduce the size of the state space problem to solve:

The first idea is to use common poles for all the material models in order to reduce the number of state variables. To achieve this goal, one need to integrate the parametric identification of the material models to the FE computation.

The second idea is to use a Ritz basis to project the nodal degrees of freedom. This leads to a reduction of the mass and stiffness operators. The choice of the basis has to be discussed in order to limit the errors induced by the projection.

The general theoretical context is first presented. The widely known constitutive equations of linear mechanics with viscoelasticity are recalled. Then the authors present a general state space formulation for such kind of problem. The size of the state space operators makes the problem very hard to compute. Thus, techniques for model reduction are presented. Theoretical aspects and numerical results are then presented.

## THEORETICAL ASPECTS

Linear viscoelasticity has been widely studied during the last decades, constitutive equations and finite element formulation are firstly recalled in order to explain the notations. Then an original state space formulation which avoid numerical scale problems is presented. Finally a new assumption is formulated; it allows to reduce the state-space problem size and consequently the computation time. All the analytical developments are illustrated with numerical results.

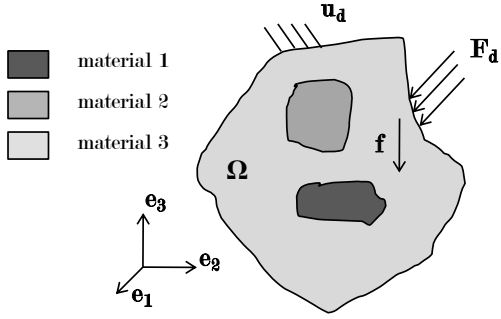

**Figure 3.** Generalized view of a heterogeneous solid with elastic and viscoelastic parts submitted to load Fd and f and kinematic boundary conditions ud

## Constitutive equations

The first Newton equation applied to an elementary volume links the acceleration to the internal stresses

$$T_{ij,j} + f_i = \rho \frac{d^2 u_i}{dt^2} \quad (1)$$

where $T_{ij}$ represents the component **ij** the Cauchy stress tensor, $f_i$ is the component **i** the volumetric loads and $u_i$ is the component i of the displacements in the (e1,e1,e3) coordinates, system, see figure 3. The compatibility equations link the displacements **u** and the strains **S**. $\rho$ is the density of the materials.

$$S_{ij,j} = \frac{1}{2}(u_{i,j} + u_{j,i}) \quad (2)$$

Linear viscoelasticity assumes that stress is a function of strain history, see for example: (E. Balmès, 2009) (R. S. Lakes, 1999) (Y. Chevalier, 2010). This translates into the existence of a relaxation function **H** given by:

$$T_{ij}(t) = \int_{-\infty}^{t} H_{ijkl}(t-\tau) S_{kl}(\tau) d\tau \quad (3)$$

The relaxation function, $H_{ijkl}$ might be different on each viscoelastic subdomain. The last equations are the boundary conditions on the displacement field **u** and the stress field **T**.

$$\begin{aligned} u &= u_d \quad \text{sur} \quad \partial\Omega_u \\ \sigma n &= F_d \quad \text{sur} \quad \partial\Omega_F \end{aligned} \quad (4)$$

Using Laplace transform, the previous equations become:

$$\begin{cases} \hat{T}_{ij,j} + \hat{f}_i = \rho \dfrac{d^2 \hat{u}_i}{dt^2} \\ \hat{S}_{ij,j} = \dfrac{1}{2}(\hat{u}_{i,j} + \hat{u}_{j,i}) \\ \hat{T}_{ij} = \hat{H}_{ijkl}(s)\hat{S}_{kl} \\ \hat{u} = \hat{u}_d \quad \text{sur} \quad \partial\Omega_u \\ \hat{\sigma} n = \hat{F}_d \quad \text{sur} \quad \Omega_F \end{cases} \quad (5)$$

In the frequency domain and where **s** is the Laplace variable.

The viscoelastic behavior of the materials might include a frequency dependence in order to correlate experimental observation. This dependence includes stiffening and frequency dependent damping. Several **rheological models** are able to take these dependences into account. The famous Kelvin-Voigt, Maxwell, and Zener models are quite efficient on a small frequency range. For the analysis on large frequency bandwidth, the Generalized Maxwell (GM) model and the fractional-derivative-based models are very accurate and realistic. Some tools had been previously developed in our team to identify viscoelastic behaviors with the GMM, see (F. Renaud, 2010) and (J.L. Dion, 1995). Thus this model is used in the present paper. Nevertheless this assumption is not essential and the following development may be used with all the viscoelastic models that are defined with rational fraction with time constants or frequency poles. In one dimension, the GM model is given by the following equation

$$\hat{T}_{11} = \left( E_\infty + \sum_i \frac{E_i \tau_i s}{\tau_i s + 1} \right) \hat{S}_{11} \quad (6)$$

Where $E_\infty$ is the long term Young Modulus (also called static modulus by some authors), $E_i$ is a dynamic modulus and $\tau_i$ is a time constant. Such kind of 1D-formulation is used to identify the behaviours of viscoelastic material with frequency dependent Storage modulus and loss factor, see figure 2 and Renaud, 2010. It is convenient to transform the previous equation in the following :

$$\hat{T}_{11} = \left( 1 + \sum_i \frac{\alpha_i \tau_i s}{\tau_i s + 1} \right) E_\infty \hat{S}_{11} \quad (7)$$

where $\alpha_i$ is the ratio $E_i/E_\infty$ which translate into the stiffening of the material according to the frequency of the excitation. The extension of the GM model to three dimensional problems leads to:

$$\hat{T} = \underbrace{\left( 1 + \sum_i \frac{\alpha_i \tau_i s}{\tau_i s + 1} \right)}_{\hat{h}(s)} H \hat{S} \quad (8)$$

Where H is the material tensor and is frequency independent. The frequency dependence is included in the function *h(s)*. The main assumption of this equation is the isotropy of the frequency dependence. In fact, even if **H** translates into anisotropy, the frequency dependence is the same for all the directions.

**Finite element approximation**

The previous problem (5) can be written in its variationnal form. This leads to the following equations when the test function **û** belongs to {**u** **C**$^1$ and **u**=**u**$_d$ for **x** **∂Ωu**}

$$\int_\Omega \rho s^2 \hat{u}_i \hat{u}_i^* dV + \int_\Omega \hat{S}_{ij} \hat{H}_{ijkl}(s,x) \hat{S}_{kl}(\hat{u}^*) dV = 0 \quad (9)$$

As the materials behaviors are frequency and space independent and linear to the displacement, the previous equations might be written separating the strain energies for each subdomain.

$$\int_\Omega \rho s^2 \hat{u}^T \hat{u}^* dV + \sum_i \hat{h}_i(s) \int_{\Omega_i} \hat{S}^T H_i \hat{S}^* dV = 0 \quad (10)$$

Where $h_i(s)$ is equal to 1 when materials are purely elastic and defined by (8) when subdomains are viscoelastic. As for a linear elastic problem, the domain can be discretized into Finite Elements and the displacement field can be interpolated by a polynomial function on each element, see for example [11]. This leads to the following algebraic equation:

$$s^2 \hat{U}^T M \hat{U}^* + \sum_i \left( \hat{h}_i(s) \hat{U}^T K_i \hat{U}^* \right) = 0 \quad (11)$$

The previous expression is transformed into the following in which the elastic terms are all assembled whereas the viscoelastic ones are kept separated. Moreover, the generalized Maxwell model is introduced.

$$s^2 \hat{U}^T M \hat{U}^* + \hat{U}^T K_e \hat{U}^* + \sum_i \left( \sum_j \frac{\alpha_{ij} \tau_{ij} s}{\tau_{ij} s + 1} \hat{U}^T K_i \hat{U}^* \right) = 0 \quad (12)$$

Where $K = \sum K_i$; $K_i$ are the stiffness matrices of each subdomain. These matrices are build using a linear finite element code introducing the long term material coefficient. If $\hat{U}^*$ belongs to {**U**=**U**$_d$ on **∂Ωu**}, the equation (12) leads to:

$$s^2 M \hat{U} + K_e \hat{U} + \sum_i \left( \sum_j \frac{\alpha_{ij} \tau_{ij} s}{\tau_{ij} s + 1} K_i \hat{U} \right) = 0 \quad (13)$$

Unfortunately, this formulation leads to very sparse matrices for which some diagonal terms are nil. For the example, figure 1, the non-zeros terms of the matrices for the material 1, 2 and 3 are plotted on the figure 4.

Due to this result, one may use separated displacement fields for each subdomain: $\hat{U}=[\hat{U}_0 \hat{U}_1 \hat{U}_2 \hat{U}_3]^T$. Where the vectors $\hat{U}_i$ are the displacement of the nodes associated the "subdomain i" and $\hat{U}_0$ to the others nodes. The equation (12) becomes:

$$s^2 \hat{U}^T M \hat{U}^* + \hat{U}^T K_e \hat{U}^* + \sum_i \left( \sum_{j=1}^n \frac{\alpha_{ij} \tau_{ij} s}{\tau_{ij} s + 1} \hat{U}_i^T K_i \hat{U}_i^* \right) = 0 \quad (14)$$

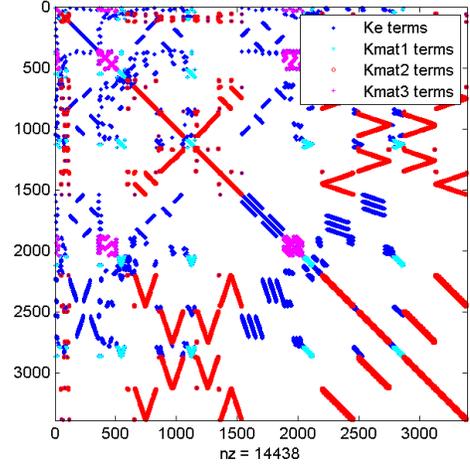

**Figure 4.** Non-zeros terms of the stiffness matrices for the different parts of the studied structure.

In this case, if the displacement fields $\hat{U}$ and $\hat{U}^*$ belong to {**U**=**U**$_d$ on **∂Ωu**}

$$s^2 M \hat{U} + K_e \hat{U} + \sum_i \left( \sum_{j=1}^n \frac{\alpha_{ij} \tau_{ij} s}{\tau_{ij} s + 1} K_i \hat{U}_i \right) = 0 \quad (15)$$

This formulation is equivalent to the (13) one but in this case the matrices are reduced to the non-zeros terms and then they are all positive and there are fewer difficulties to compute their inverse.

**State space formulation**

Equation (13) is not in a classical form in order to be solved. Two ways can be considered to solve such kind of equation : the first one is to use a specific algorithm to extract the roots and associated vectors, see for example [9]; the second one is to transform the problem is order to obtain a generalized eigenvalue problem. This has been done in several papers before: for example [2] introduce internal state in their formulation to finally obtain such kind of generalized eigenvalue problem.

The first internal state that have been chosen is $\hat{U}_i^{(j)}=s\hat{U}_i^{(j-1)}$, see [8]. In this case the operators built in the state-space are very badly scaled; It can lead to erroneous results when the number of states is great. The other form that has been proposed is widely influenced by the one proposed by [2]:

$$\hat{U}_i^{(1)} = s \hat{U}_i \text{ and } \hat{U}_i^{(j+1)} = \frac{\tau_{ij}}{\tau_{ij} s + 1} \hat{U}_i^{(1)}$$

$$sM \hat{U}^{(1)} + K_e \hat{U} + \sum_i \left( \sum_{j=1}^n \alpha_{ij} K_i \hat{U}_i^{(j+1)} \right) = 0 \quad (16)$$

The state-space formulation obtained with (16) is :

$$s\begin{bmatrix} I & 0 & 0 & 0 & 0 \\ 0 & M & 0 & 0 & 0 \\ 0 & 0 & T_1 & 0 & 0 \\ 0 & 0 & 0 & \ddots & 0 \\ 0 & 0 & 0 & 0 & T_m \end{bmatrix}\begin{bmatrix} \hat{U} \\ \hat{U}^{(1)} \\ \hat{U}_1^{(2\cdots n_1)} \\ \vdots \\ \hat{U}_m^{(2\cdots n_m)} \end{bmatrix} =$$

$$\begin{bmatrix} 0 & I & 0 & 0 & 0 \\ -K & 0 & -A_1 & \cdots & -A_m \\ 0 & d(T_1) & -I & 0 & 0 \\ 0 & \vdots & 0 & \ddots & 0 \\ 0 & d(T_m) & 0 & 0 & -I \end{bmatrix}\begin{bmatrix} \hat{U} \\ \hat{U}^{(1)} \\ \hat{U}_1^{(2\cdots n_1)} \\ \vdots \\ \hat{U}_m^{(2\cdots n_m)} \end{bmatrix} \quad (17)$$

Where

$\hat{U} = [\hat{U}_0\, \hat{U}_1\, \ldots\, \hat{U}_m]^T$

$\hat{U}^{(1)} = [\hat{U}_0^{(1)}\, \hat{U}_1^{(1)}\, \ldots\, \hat{U}_m^{(1)}]^T$

$\hat{U}_1^{(2\ldots n_1)} = [\hat{U}_1^{(2)}\, \ldots\, \hat{U}_1^{(n_1)}]^T$

$\hat{U}_m^{(2\ldots n_m)} = [\hat{U}_m^{(2)}\, \ldots\, \hat{U}_m^{(n_m)}]^T$

and $d(T_i) = \begin{bmatrix} \tau_{11}I & 0 & 0 \\ \vdots & 0 & 0 \\ \tau_{1n_1}I & 0 & 0 \\ \vdots & \ddots & \vdots \\ 0 & 0 & \tau_{m1}I \\ 0 & 0 & \vdots \\ 0 & 0 & \tau_{mn_m}I \end{bmatrix}$

$T_i = \begin{bmatrix} \tau_{i1}I & 0 & 0 \\ 0 & \ddots & 0 \\ 0 & 0 & \tau_{in_i}I \end{bmatrix}$ and $A_i = \begin{bmatrix} \alpha_{i1}K_i & 0 & 0 \\ 0 & \ddots & 0 \\ 0 & 0 & \alpha_{in_i}K_i \end{bmatrix}$

Equation (17) is very close to the formulation of [2].

## MODAL REDUCTION

The main problem with the last is the size of the state space problem. If the number of terms in each Generalized Maxwell zeries in equal to n, the size of the state-space operators is equal to $(n+1)n_{DOF}$ where nDOF is the number of degrees of freedom of the initial finite element model. [10]proposed a state space formulation in which the mass and stiffness operators **M** and **K**, **$K_i$** are reduced through a projection in a suitable truncated modal basis.

### Simple Formulation

The first idea is to use the modal basis $\Psi_\infty$ that have been obtained for the conservative problem, i.e. considering viscoelastic parts as elastic ones and taking the long term moduli into account for the calculus, see figure 3.

$$(s^2 M + K(E, E_{i\infty}))\psi_\infty = 0 \quad (18)$$

Considering this, $\hat{U} = \Psi_\infty Q$. Under this assumption, the previous formulation becomes:

$$Q^{*T}\left[s^2 \underbrace{\Psi_\infty^T M \Psi_\infty}_{m_\infty} + \underbrace{\Psi_\infty^T K \Psi_\infty}_{k_\infty} + \sum_i\left(\sum_j \frac{\alpha_{ij}\tau_{ij}s}{\tau_{ij}s+1}\underbrace{\Psi_\infty^T K_i \Psi_\infty}_{k_{i\infty}}\right)\right]Q = 0$$

Due to the projection, the sparse matrices **$K_i$** become full and the sizes of **$m_\infty$**, **$k_\infty$** and **$k_{i\infty}$** are identical and equal to **$N_{\Psi\infty}$** the number of eigenvectors that have been retained for the projection. Thus the state variables are simplified:

$$\hat{Q}^{(1)} = s\hat{Q}$$

$$\hat{Q}_i^{(j+1)} = \frac{\tau_{ij}}{\tau_{ij}s+1}\hat{Q}^{(1)} \quad (20)$$

$$sm_\infty \hat{Q}^{(1)} + k_\infty \hat{Q} + \sum_i\left(\sum_{j=1}^n \alpha_{ij}k_{i\infty}\hat{Q}_i^{(j+1)}\right) = 0$$

This leads to the following state-space formulation:

$$s\begin{bmatrix} I & 0 & 0 & 0 & 0 \\ 0 & m_\infty & 0 & 0 & 0 \\ 0 & 0 & t_1 & 0 & 0 \\ 0 & 0 & 0 & \ddots & 0 \\ 0 & 0 & 0 & 0 & t_m \end{bmatrix}\begin{bmatrix} \hat{Q} \\ \hat{Q}^{(1)} \\ \hat{Q}^{(2\cdots n_1)} \\ \vdots \\ \hat{Q}^{(2\cdots n_m)} \end{bmatrix} =$$

$$\begin{bmatrix} 0 & I & 0 & 0 & 0 \\ -k_\infty & 0 & -a_1 & \cdots & -a_m \\ 0 & d(t_1) & -I & 0 & 0 \\ 0 & \vdots & 0 & \ddots & 0 \\ 0 & d(t_m) & 0 & 0 & -I \end{bmatrix}\begin{bmatrix} \hat{Q} \\ \hat{Q}^{(1)} \\ \hat{Q}^{(2\cdots n_1)} \\ \vdots \\ \hat{Q}^{(2\cdots n_m)} \end{bmatrix} \quad (13)$$

With

$t_i = \begin{bmatrix} \tau_{i1}I & 0 & 0 \\ 0 & \ddots & 0 \\ 0 & 0 & \tau_{in_i}I \end{bmatrix}$ $d(t_i) = \begin{bmatrix} \tau_{11}I \\ \vdots \\ \tau_{1n_1}I \\ \vdots \\ \tau_{m1}I \\ \vdots \\ \tau_{mn_m}I \end{bmatrix}$ $a_i = [\alpha_{i1}k_{i\infty}\ \cdots\ \alpha_{in_i}k_{i\infty}]$

The size of the state space problem is then reduced to $N_{\Psi\infty}(2+(n-1)m)$ where m is the number of viscoelastic materials.

Résultats sur la poutre

### Common Polesassumption

In order to reduce the computation cost, the number of state may be reduced choosing "common poles" or "common relaxation time" in our identification. Figure 1 shows the Frequency dependence of the three viscoelastic materials identified separately and using the method published by [6] and

[5]. The parameters that have been used are summarized in the next table:

**Table 1 – Numerical properties of the three viscoleastic materials described in Figure 2.**

| Material 1 | | Material 2 | | Material 3 | |
|---|---|---|---|---|---|
| $\alpha_{1j}$ | $\tau_{1j}$ | $\alpha_{2j}$ | $\tau_{2j}$ | $\alpha_{3j}$ | $\tau_{3j}$ |
| 8.86E-01 | 1.88E-01 | 6.37E-01 | 1.66E-01 | 5.70E-01 | 2.32E-01 |
| 1.63E+00 | 2.28E-02 | 1.04E+00 | 1.34E-02 | 8.92E-01 | 3.58E-02 |
| 3.24E+00 | 2.23E-03 | 1.85E+00 | 1.22E-03 | 1.45E+00 | 3.65E-03 |
| 5.69E+00 | 2.21E-04 | 2.84E+00 | 1.26E-04 | 2.18E+00 | 3.43E-04 |
| 1.39E+01 | 2.59E-05 | 6.31E+00 | 2.36E-05 | 3.78E+00 | 3.20E-05 |
| E∞ | | E∞ | | E∞ | |
| 6.15E+06 | Pa | 9.60E+06 | Pa | 4.74E+06 | Pa |

Using "Common Pôles" allows reducing the number of state variables and it is possible to constrain the identification process to obtain the material properties simultaneously, see figure 7. These properties are summarized in Table 3

**Table 3 – Numerical properties of the three viscoleastic materials described in Figure 7.**

| $\tau_j$ | $\alpha_{1j}$ | $\alpha_{2j}$ | $\alpha_{2j}$ |
|---|---|---|---|
| **2.66E-02** | 4.656E-01 | 3.119E-01 | 4.215E-01 |
| **5.97E-03** | 1.211E+00 | 6.220E-01 | 7.940E-01 |
| **6.16E-04** | 2.461E+00 | 1.230E+00 | 1.316E+00 |
| **5.29E-05** | 3.858E+00 | 1.850E+00 | 2.040E+00 |
| **2.41E-05** | 4.885E+00 | 2.365E+00 | 1.071E+00 |
| | E∞ | E∞ | E∞ |
| | 7.37E+06 Pa | 2.18E+07 Pa | 5.12E+06 Pa |

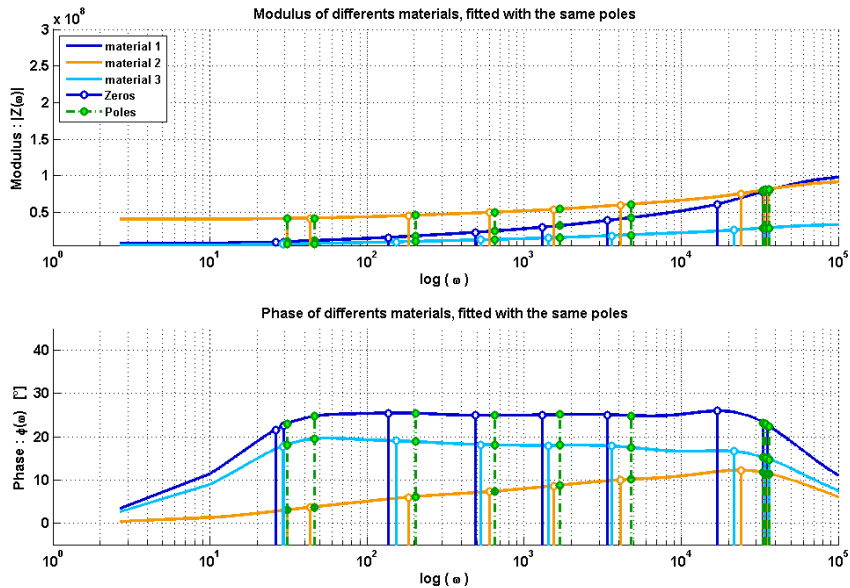

**Figure 5.** Frequency dependence of the three viscoelastic materials identified simultaneously

Due to this assumption the state variable can be defined as:

$$\hat{Q}^{(1)} = s\hat{Q}$$

$$\hat{Q}^{(j+1)} = \frac{\tau_j}{\tau_j s + 1} \hat{Q}^{(1)} \quad (21)$$

$$sm_\infty \hat{Q}^{(1)} + k_\infty \hat{Q} + \sum_i \left( \sum_{j=1}^n \alpha_{ij} k_{i\infty} \right) \hat{Q}^{(j+1)} = 0$$

This leads to the following state space formulation:

$$s \begin{bmatrix} I & 0 & 0 \\ 0 & m_\infty & 0 \\ 0 & 0 & t \end{bmatrix} \begin{bmatrix} \hat{Q} \\ \hat{Q}^{(1)} \\ \hat{Q}^{(2\cdots n)} \end{bmatrix} = \begin{bmatrix} 0 & I & 0 \\ -k_\infty & 0 & -a \\ 0 & d(t) & -I \end{bmatrix} \begin{bmatrix} \hat{Q} \\ \hat{Q}^{(1)} \\ \hat{Q}^{(2\cdots n)} \end{bmatrix} \quad (22)$$

With

$$t = \begin{bmatrix} \tau_1 I & 0 & 0 \\ 0 & \ddots & 0 \\ 0 & 0 & \tau_n I \end{bmatrix} \quad d(t) = \begin{bmatrix} \tau_1 I \\ \vdots \\ \tau_n I \end{bmatrix} \text{ and}$$

$$a = \begin{bmatrix} \sum \alpha_{i1} k_{i\infty} & \cdots & \sum \alpha_{in} k_{i\infty} \end{bmatrix}$$

The size of the state space problem is then reduced to **$N_{\Psi_\infty}(2+(n-1))$.**

### Projection errors

Due to the reduction through a projection in a suitable truncated modal basis, the complex modes and eigenvalues that have been identified are erroneous. In order to quantify the error, on can compute the residue, including the complex eigenvector and eigenvalue in the equation (13) where the displacement $\hat{U} = \Psi_\infty \phi_i$. Therefore the error for each complex eigenvector is defined by:

$$s_i^2 M \Psi_\infty \phi_i + K \Psi_\infty \phi_i + \sum_i \left( \sum_j \frac{\alpha_{ij} \tau_{ij} s_i}{\tau_{ij} s_i + 1} K_i \Psi_\infty \phi_i \right) = \varepsilon_i \quad (23)$$

### Enhancement of the projection basis

In order to reduce the errors on each eigenmodes, one can enhance the projection basis by iterative computation on the projection error (23), see [12] or [13]. Nevertheless these methods are computationally expensive. Here we propose to enhance the basis using the modal basis build with a high frequency modulus:

$$\left( s^2 M + K\left( E, E_{iHF} \right) \right) \psi_{HF} = 0 \quad (24)$$

Considering this, $\hat{U}=[\Psi_\infty \Psi_{HF}]Q=\hat{U}=[\Psi_{Enh}]Q$ Under this assumption, the previous formulation becomes:

$$Q^{*T} \begin{bmatrix} s^2 \underbrace{\Psi_{Enh}^T M \Psi_{Enh}}_{m_{Enh}} + \underbrace{\Psi_{Enh}^T K \Psi_{Enh}}_{k_{Enh}} + \\ \sum_i \left( \sum_j \frac{\alpha_{ij} \tau_{ij} s}{\tau_{ij} s + 1} \underbrace{\Psi_{Enh}^T K_i \Psi_{Enh}}_{k_{iEnh}} \right) \end{bmatrix} Q = 0 \quad (25)$$

The error is then calculated using the formula (23).

## CONCLUSIONS

Following the objectives that have been previously introduced, the paper presents ideas to reduce the computational cost of CEA with viscoelastic materials. The analytical developments presented in this paper allow reducing the number of internal states. Moreover the use of an enhanced modal Ritz-basis allows to reduce the errors induced by a projection on a real basis.

## ACKNOWLEDGMENTS

The authors wish to thanks Thierry Pasquet and RemiLemaire from the CSB division,NVH department of Robert Bosch France for their technical and financial help.